\documentclass{article}
\usepackage{epsfig}
\usepackage{amsmath}
\usepackage{amsfonts}
\usepackage{amssymb}
\usepackage{euscript}
\usepackage{graphicx}
\usepackage{multicol}
\usepackage{lscape}
\usepackage{pstricks}
\usepackage{multirow}

\begin{document}
\sloppy

\begin{center}
{{\Huge\bfseries  The relations between main stellar parameters\\}}
\end{center}

\begin{center}
{\Large

\itshape{ B.V.Vasiliev}}
\end{center}

The theoretical consideration of the stellar substance equilibrium  \cite{vas} allows to conclude that main stellar parameters are connected by the relation:

\begin{equation}
{\mathbb{T}}_0 \mathbb{R}_0=Const\cdot \mathbb{M}^{5/4}.\label{a}
\end{equation}

Where $\mathbb{M}$,$\mathbb{R}_0$ and $\mathbb{T}_0$ are a mass of the star, its radius and temperature on its surface.

If to take into account the thermodynamic characteristics of stellar substance \cite{vas},
 one can see that this relation is effect of two other correlations:

\begin{equation}
\frac{\mathbb{M}^2}{\mathbb{R}_0^3}=const\label{rm}
\end{equation}

and

\begin{equation}
\mathbb{T}_0=const\cdot\mathbb{M}^{7/12}\label{tm}.
\end{equation}

If main parameters are expressed through corresponding solar values
$\tau\equiv\frac{\mathbb{T}_0}{\mathbb{T}_\odot}$,$\rho\equiv\frac{\mathbb{R}_0}{\mathbb{R}_\odot}$
 and $\mu\equiv\frac{\mathbb{M}}{\mathbb{M}_\odot}$, that Eqs.({\ref{a}}), ({\ref{rm}}) and ({\ref{tm}}) can be rewritten as
 \begin{equation}
\frac{\tau \rho}{\mu^{5/4}}=1\label{mrt2},
\end{equation}

\begin{equation}
\frac{\rho}{\mu^{2/3}}=1\label{rm2}
\end{equation}

and

\begin{equation}
\frac{\tau}{\mu^{7/12}}=1\label{tm2}.
\end{equation}

Simultaneously masses, radiuses and surface temperatures can be measured for close binary stars. The measuring data of these parameters are gathered in consolidated table in dissertation (in Russian) \cite{Kh}. As a matter of convenience of readers, this table together with references on original studies is quoted \cite{vas}.
On base of this  data, relations (\ref{mrt2}), (\ref{rm2}) and (\ref{tm2}) are gathered in the Table. 

    The analysis of these data leads to few conclusions.
The averaging over all tabulated stars gives
\begin{equation}
<\frac{\tau}{\mu^{7/12}}>=1.007\pm 0.07 .
\end{equation}
and we can conclude that  the variability of measured data of surface temperatures and stellar masses has statistical character. Secondly, Eq.({\ref{tm2}}) is valid for all hot stars (exactly for all stars which are gathered in Table).

The problem with the averaging of $\frac{\rho}{\mu^{2/3}}$ looks different. There are a few of giants and super-giants in this Table.  The values of ratio  $\frac{\rho}{\mu^{2/3}}$  are more than 2 for them. It seems that, if to  exclude these stars  from consideration,  the averaging over stars of the main sequence gives value close to 1. Evidently, it needs in more detail consideration.

\begin{table}

{\hspace{-4.5cm}Table.

\hspace{-4.5cm}The relations between main stellar parameters}

{
%\scriptsize
\hspace{-4.5cm}
\tiny
\begin{tabular}
{||r|l|r|r|r|r|r|r|r||}\hline \hline
   & & & & & & & &  \\
N & {Star} & &$\mu\equiv\frac{\mathbb{M}}{\mathbb{M}_{\odot}}$
&$\rho\equiv\frac{\mathbb{R}_0}{\mathbb{R}_{\odot}}$
&$\tau\equiv\frac{\mathbb{T}_0}{\mathbb{T}_{\odot}}$
&$\frac{\rho}{\mu^{2/3}}$ & $\frac{\tau}{\mu^{7/12}}$
&$\frac{\rho\tau}{\mu^{5/4}}$
\\[0.4cm]\hline
& & & & & & & &  \\
&  & 1 & 1.48 & 1.803 & 1.043 & 1.38 & 0.83 & 1.15 \\[0.2cm]  \cline{3-9}
1&   BW Aqr  &&&&&&&\\
&  & 2 & 1.38 & 2.075 & 1.026 & 1.67 & 0.85  & 1.42 \\
& & & & & & & &  \\
\hline
& & & & & & & &  \\
&  & 1 & 2.4  & 2.028 & 1.692 & 1.13  & 1.01 & 1.15 \\[0.2cm]  \cline{3-9}
2 & V 889 Aql &       &  &&     &  &  &  \\
& &  2 & 2.2  & 1.826 & 1.607 &  1.08 & 1.01 & 1.09 \\
& & & & & & & &  \\
\hline
 & & & & & & & &  \\
& &1 & 6.24   & 4.512 & 3.043 & 1.33 & 1.04 & 1.39 \\[0.2cm]  \cline{3-9}
3 & V 539 Ara &       &   &&    &  &  &  \\
& & 2& 5.31   & 4.512 & 3.043 &  1.12 & 1.09  & 1.23 \\
 & & & & & & & &  \\
\hline
 & & & & & & & &  \\
& &1 & 3.31   & 2.58  & 1.966 & 1.16  & 0.98 & 1.13 \\[0.2cm]  \cline{3-9}
  4 & AS Cam  &       &     &&  &  &  &  \\
& & 2& 2.51   &1.912  & 1.709 & 1.03 & 1.0 & 1.03 \\
 & & & & & & & &  \\
\hline
 & & & & & & & &  \\
& & 1 &  22.8 & 9.35  & 5.658 & 1.16 & 0.91 & 1.06 \\[0.2cm]  \cline{3-9}
5 & EM Car  &       &       & && &  &  \\
& & 2 & 21.4  & 8.348 & 5.538 & 1.08 & 0.93 & 1.00\\
 & & & & & & & &  \\
\hline
 & & & & & & & &  \\
& &1 & 13.5  & 4.998 & 5.538 & 0.88  & 1.08 & 0.95 \\[0.2cm]  \cline{3-9}
  6 & GL Car  &&  &&&&&\\
& & 2& 13 & 4.726 & 4.923 & 0.85 & 1.1 & 0.94  \\
 & & & & & & & &  \\
\hline
 & & & & & & & &  \\
& &1 & 9.27 & 4.292 & 4 & 0.97 & 1.09 & 1.06  \\[0.2cm]  \cline{3-9}
  7 & QX Car    &&&&&&&\\
& & 2& 8.48 & 4.054 & 3.829 & 0.975 & 1.1 & 1.07 \\
 & & & & & & & &  \\
\hline
 & & & & & & & &  \\
& &1 & 6.7 & 4.591 & 3.111 & 1.29 & 1.02& 1.32\\[0.2cm]  \cline{3-9}
  8 & AR Cas &&&&&&&\\
& & 2&  1.9 & 1.808 & 1.487 & 1.18 & 1.02& 1.21 \\
 & & & & & & & &  \\
\hline
 & & & & & & & &  \\
& &1 &  1.4 & 1.616 & 1.102 & 1.29 & 0.91 & 1.17 \\[0.2cm]  \cline{3-9}
  9 & IT Cas    &&&&&&&\\
& & 2&  1.4 & 1.644 & 1.094 & 1.31 & 0.90& 1.18 \\
 & & & & & & & &  \\
\hline
 & & & & & & & &  \\
& &1 &  7.2 & 4.69 & 4.068 & 1.25 & 1.29 & 1.62 \\[0.2cm]  \cline{3-9}
  10 & OX Cas   &&&&&&&\\
& & 2&  6.3 & 4.54 & 3.93 & 1.33 & 1.34 & 1.79  \\
 & & & & & & & &  \\
\hline
 & & & & & & & &  \\
& &1 &  2.79 & 2.264 & 1.914 & 1.14 & 1.05 & 1.20\\[0.2cm]  \cline{3-9}
  11 & PV Cas   &&&&&&&\\
& & 2 &  2.79 & 2.264 & 2.769 & 1.14 & 1.05 & 1.20 \\
 & & & & & & & &  \\
\hline
 & & & & & & & &  \\
& &1 & 5.3 & 4.028 & 2.769  & 1.32 & 1.05 & 1.39 \\[0.2cm]  \cline{3-9}
  12 & KT Cen   &&&&&&&\\
& & 2 & 5 & 3.745 & 2.701 & 1.28 & 1.06 & 1.35 \\
 & & & & & & & &  \\
\hline
 & & & & & & & &  \\
& &1 &  11.8 & 8.26 & 4.05 & 1.59 & 0.96 & 1.53 \\[0.2cm]  \cline{3-9}
  13 & V 346 Cen &&&&&&&\\
& & 2 & 8.4 & 4.19 & 3.83 & 1.01 & 1.11 & 1.12 \\
 & & & & & & & &  \\
\hline
 & & & & & & & &  \\
& & 1 &  11.8 & 8.263 & 4.051 & 1.04 & 1.06 & 1.11 \\[0.2cm]  \cline{3-9}
  14 & CW Cep   &&&&&&&\\
& & 2 & 11.1 & 4.954 & 4.393 & 1.0 & 1.08 & 1.07 \\
 & & & & & & & &  \\
\hline
 & & & & & & & &  \\
& & 1 & 2.02 & 1.574 & 1.709  & 0.98 & 1.13 & 1.12 \\[0.2cm]  \cline{3-9}
  15 & EK Cep   &&&&&&&\\
& & 2 &  1.12 & 1.332 & 1.094   & 1.23 & 1.02 & 1.26 \\
 & & & & & & & &  \\
\hline
 & & & & & & & &  \\
& & 1 & 2.58 & 3.314 & 1.555  & 1.76 & 0.89 & 1.57 \\[0.2cm]  \cline{3-9}
  16 & $\alpha$ Cr B &&&&&&&\\
& & 2 &  0.92 & 0.955 & 0.923 & 1.01 & 0.97 & 0.98 \\
 & & & & & & & &  \\
\hline
 & & & & & & & &  \\
& & 1 & 17.5 & 6.022 & 5.66 & 0.89 & 1.06 &  0.95 \\[0.2cm]  \cline{3-9}
  17 & Y Cyg    &&&&&&&\\
& & 2 & 17.3 & 5.68 & 5.54 & 0.85 & 1.05 & 0.89 \\
 & & & & & & & &  \\
\hline\hline
 %& & & & & & & &  \\

\end{tabular}}\label{mrt}
\end{table}

\clearpage

\begin{table}

{\hspace{-4.5cm}Table(continuation).

\hspace{-4.5cm}The relations between main stellar parameters
}

{
\tiny
{
%\scriptsize
\hspace{-4.5cm}
\begin{tabular}
 {||r|l|r|r|r|r|r|r|r||}\hline
\hline
   & & & & & & & &  \\
 N & \tiny{Star} & n%component
 &$\mu\equiv\frac{\mathbb{M}}{\mathbb{M}_{\odot}}$ &
$\rho\equiv\frac{\mathbb{R}_0}{\mathbb{R}_{\odot}}$ &
$\tau\equiv\frac{\mathbb{T}_0}{\mathbb{T}_{\odot}}$ &
 $\frac{\rho}{\mu^{2/3}}$& $\frac{\tau}{\mu^{7/12}}$ & $\frac{\rho\tau}{\mu^{5/4}}$
\\[0.4cm]\hline\hline
& & & & & & & &  \\
& &1 &  14.3 & 17.08 & 3.54  & 2.89 & 0.75 & 2.17 \\[0.2cm]  \cline{3-9}
  18 & Y 380 Cyg &&&&&&&\\
& & 2&  8 & 4.3 & 3.69 & 1.07& 1.1& 1.18 \\
 & & & & & & & &  \\
 \hline
 & & & & & & & &  \\
& & 1 &  14.5 & 8.607 & 4.55 & 1.45 & 0.95 &  1.38 \\[0.2cm]  \cline{3-9}
  19 & V 453 Cyg &&&&&&&\\
& & 2 &  11.3 & 5.41 & 4.44 & 1.07 & 1.08 & 1.16 \\
 & & & & & & & &  \\
\hline
 & & & & & & & &  \\
& & 1 & 1.79 & 1.567 & 1.46 & 1.06 & 1.04 & 1.11 \\[0.2cm]  \cline{3-9}
  20 & V 477 Cyg &&&&&&&\\
& & 2 & 1.35 & 1.27 & 1.11& 1.04 & 0.93 & 0.97 \\
 & & & & & & & &  \\
\hline
 & & & & & & & &  \\
& &1 & 16.3 & 7.42 & 5.09& 1.15 & 1.0 & 1.15 \\[0.2cm]  \cline{3-9}
  21 & V 478 Cyg &&&&&&&\\
& & 2 & 16.6 & 7.42 & 5.09 & 1.14 & 0.99 & 1.13 \\
 & & & & & & & &  \\
\hline
 & & & & & & & &  \\
& & 1 & 2.69 & 2.013 & 1.86 & 1.04 & 1.05 & 1.09 \\[0.2cm]  \cline{3-9}
  22 & V 541 Cyg &&&&&&&\\
& & 2 & 2.6 & 1.9 & 1.85 & 1.0 & 1.6 & 1.06 \\
 & & & & & & & &  \\
\hline
 & & & & & & & &  \\
& &1 & 1.39 & 1.44 & 1.11 & 1.16 & 0.92 & 0.92 \\[0.2cm]  \cline{3-9}
  23 & V 1143 Cyg &&&&&&&\\
& & 2 &  1.35 & 1.23 & 1.09 & 1.0 & 0.91 & 0.92 \\
 & & & & & & & &  \\
\hline
 & & & & & & & &  \\
& &1 & 23.5 & 19.96 & 4.39 & 2.43 & 0.67 & 1.69 \\[0.2cm]  \cline{3-9}
  24 & V 1765 Cyg &&&&&&&\\
& & 2& 11.7 & 6.52 & 4.29 & 1.26 & 1.02 & 1.29 \\
 & & & & & & & &  \\
\hline
 & & & & & & & &  \\
& &1 & 5.15 & 2.48 & 2.91 & 0.83 & 1.12 & 0.93 \\[0.2cm]  \cline{3-9}
  25 & DI Her    &&&&&&&\\
& & 2& 4.52 & 2.69 & 2.58 & 0.98 & 1.07 & 1.05 \\
 & & & & & & & &  \\
\hline
 & & & & & & & &  \\
 & & & & & & & &  \\
&  & 1 & 4.25 & 2.71 & 2.61     &    1.03 & 1.12 & 1.16\\[0.2cm]\cline{3-9}
26 &   HS Her  &       &     &&  &  & &  \\
&  & 2 & 1.49 & 1.48 & 1.32 & 1.14 & 1.04  & 1.19 \\
& & & & & & & &  \\
\hline
& & & & & & & &  \\
&  & 1 & 3.13  & 2.53 & 1.95 & 1.18  & 1.00 & 1.12 \\[0.2cm]  \cline{3-9}
27 & CO Lac &       &      && &  &  &  \\
& &  2 & 2.75  & 2.13 & 1.86 &  1.08 & 1.01 & 1.09 \\
& & & & & & & &  \\
\hline
& & & & & & & &  \\
& &1 & 6.24   & 4.12 & 2.64 & 1.03 & 1.08 & 1.11 \\[0.2cm]  \cline{3-9}
28 & GG Lup  &       &   &&    &  &  &  \\
& & 2& 2.51   & 1.92 & 1.79 &  1.04 & 1.05  & 1.09 \\
& & & & & & & &  \\
\hline
& & & & & & & &  \\
& &1 & 3.6   & 2.55 & 2.20 & 1.09  & 1.04 & 1.14 \\[0.2cm]  \cline{3-9}
29  & RU Mon  &       &&&       &  &  &  \\
& & 2& 3.33   & 2.29  & 2.15 & 1.03 & 1.07 & 1.10 \\
& & & & & & & &  \\
\hline
& & & & & & & &  \\
& & 1 & 2.5 & 4.59  & 1.33 & 2.49 & 0.78 & 1.95 \\[0.2cm]  \cline{3-9}
30 & GN Nor  &       &&&       &  &  &  \\
& & 2 & 2.5  & 4.59 & 1.33 & 2.49 & 0.78 & 1.95\\
& & & & & & & &  \\
\hline
& & & & & & & &  \\
& &1 & 5.02 & 3.31 & 2.80 & 1.13  & 1.09 & 1.23 \\[0.2cm]  \cline{3-9}
31 & U Oph    &&&&&&&\\
& & 2& 4.52 & 3.11 & 2.60 & 1.14 & 1.08 & 1.23  \\
& & & & & & & &  \\
\hline
& & & & & & & &  \\
& &1 & 2.77 & 2.54 & 1.86 & 1.29 & 1.03 & 1.32  \\[0.2cm]  \cline{3-9}
32 & V 451 Oph    &&&&&&&\\
& & 2& 2.35 & 1.86 & 1.67 & 1.05 & 1.02 & 1.07 \\
& & & & & & & &  \\
\hline
& & & & & & & &  \\
& &1 & 19.8 & 14.16 & 4.55 & 1.93 & 0.80 & 1.54 \\[0.2cm]  \cline{3-9}
33 & $\beta$ Ori &&&&&&&\\
& & 2&  7.5 & 8.07 & 3.04 & 2.11 & 0.94 & 1.98 \\
& & & & & & & &  \\
\hline
& & & & & & & &  \\
& &1 &  2.5 & 1.89 & 1.81 & 1.03 & 1.06 & 1.09 \\[0.2cm]  \cline{3-9}
34 & FT Ori    &&&&&&&\\
& & 2&  2.3 & 1.80 & 1.62 & 1.03 & 1.0 & 1.03 \\
& & & & & & & &  \\
\hline\hline
%& & & & & & & &  \\

\end{tabular}}}\label{mrt}
\end{table}

\clearpage

\begin{table}

{\hspace{-4.5cm}Table(continuation).

\hspace{-4.5cm}The relations between main stellar parameters
}

{
\tiny
{
%\scriptsize
\hspace{-4.5cm}
\begin{tabular}
 {||r|l|r|r|r|r|r|r|r||}\hline
\hline
   & & & & & & & &  \\
 N & \tiny{Star} & n%component
 &$\mu\equiv\frac{\mathbb{M}}{\mathbb{M}_{\odot}}$ &
$\rho\equiv\frac{\mathbb{R}_0}{\mathbb{R}_{\odot}}$ &
$\tau\equiv\frac{\mathbb{T}_0}{\mathbb{T}_{\odot}}$ &
 $\frac{\rho}{\mu^{2/3}}$& $\frac{\tau}{\mu^{7/12}}$ & $\frac{\rho\tau}{\mu^{5/4}}$
\\[0.4cm]\hline\hline
& & & & & & & &  \\
& &1 &  5.36 & 3.0 & 2.91 & 0.98 & 1.09 & 1.06 \\[0.2cm]  \cline{3-9}
35 & AG Per   &&&&&&&\\
& & 2&  4.9 & 2.61 & 2.91 & 0.90 & 1.15 & 1.04  \\
& & & & & & & &  \\
\hline
& & & & & & & &  \\
& &1 &  3.51 & 2.44 & 2.27 & 1.06 & 1.09 & 1.16\\[0.2cm]  \cline{3-9}
36 & IQ Per   &&&&&&&\\
& & 2 &  1.73 & 1.50 & 2.27 & 1.04 & 1.00 & 1.05 \\
& & & & & & & &  \\
\hline
& & & & & & & &  \\
& &1 & 3.93 & 2.85 & 2.41  & 1.14 & 1.08 & 1.24 \\[0.2cm]  \cline{3-9}
37 & $\varsigma$ Phe   &&&&&&&\\
& & 2 & 2.55 & 1.85 & 1.79 & 0.99 & 1.04 & 1.03 \\
& & & & & & & &  \\
\hline
& & & & & & & &  \\
& &1 &  2.5 & 2.33 & 1.74 & 1.27 & 1.02 & 1.29 \\[0.2cm]  \cline{3-9}
38 & KX Pup &&&&&&&\\
& & 2 & 1.8 & 1.59 & 1.38 & 1.08 & 0.98 & 1.06 \\
& & & & & & & &  \\
\hline
& & & & & & & &  \\
& & 1 &  2.88 & 2.03 & 1.95 & 1.00 & 1.05 & 1.05 \\[0.2cm]  \cline{3-9}
39 & NO Pup   &&&&&&&\\
& & 2 & 1.5 & 1.42 & 1.20 & 1.08 & 0.94 & 1.02 \\
& & & & & & & &  \\
\hline
& & & & & & & &  \\
& & 1 & 2.1 & 2.17 & 1.49  & 1.32 & 0.96 & 1.27 \\[0.2cm]  \cline{3-9}
40 & VV Pyx   &&&&&&&\\
& & 2 &  2.1 & 2.17 & 1.49   & 1.32 & 0.96 & 1.27 \\
& & & & & & & &  \\
\hline
& & & & & & & &  \\
& & 1 & 2.36 & 2.20 & 1.59  & 1.24 & 0.96 & 1.19 \\[0.2cm]  \cline{3-9}
41 & YY Sgr &&&&&&&\\
& & 2 &  2.29 & 1.99 & 1.59 & 1.15 & 0.98 & 1.12 \\
& & & & & & & &  \\
\hline
& & & & & & & &  \\
& & 1 & 2.1 & 2.67 & 1.42 & 1.63 & 0.92 &  1.50 \\[0.2cm]  \cline{3-9}
42 & V 523 Sgr   && &&&&&\\
& & 2 & 1.9 & 1.84 & 1.42 & 1.20 & 0.98 & 1.17 \\
& & & & & & & &  \\
\hline
& & & & & & & &  \\
& &1 & 2.11 & 1.9 & 1.30 & 1.15 & 0.84 & 0.97 \\[0.2cm]  \cline{3-9}
43 & V 526 Sgr &&&&&&&\\
& & 2&  1.66 & 1.60 & 1.30 & 1.14 & 0.97 & 1.10 \\
& & & & & & & &  \\
\hline
& & & & & & & &  \\
& & 1 &  2.19 & 1.83 & 1.52 & 1.09 & 0.96 &  1.05  \\[0.2cm]  \cline{3-9}
44 & V 1647 Sgr &&&&&&&\\
& & 2 &  1.97 & 1.67 & 4.44 & 1.06 & 1.02 & 1.09 \\
& & & & & & & &  \\
\hline
& & & & & & & &  \\
& & 1 & 3.0 & 1.96 & 1.67 & 0.94 & 0.88 & 0.83 \\[0.2cm]  \cline{3-9}
45 & V 2283 Sgr &&&&&&&\\
& & 2 & 2.22 & 1.66 & 1.67 & 0.97 & 1.05 & 1.02 \\
& & & & & & & &  \\
\hline
& & & & & & & &  \\
& &1 & 4.98 & 3.02 & 2.70 & 1.03 & 1.06 & 1.09 \\[0.2cm]  \cline{3-9}
46 & V 760 Sco &&&&&&&\\
& & 2 & 4.62 & 2.64 & 2.70 & 0.95 & 1.11 & 1.05 \\
& & & & & & & &  \\
\hline
& & & & & & & &  \\
& & 1 & 3.2 & 2.62 & 1.83 & 1.21 & 0.93 & 1.12 \\[0.2cm]  \cline{3-9}
47 & AO Vel &&&&&&&\\
& & 2 & 2.9 & 2.95 & 1.83 & 1.45 & 0.98 & 1.43 \\
& & & & & & & &  \\
\hline
& & & & & & & &  \\
& &1 & 3.21 & 3.14 & 1.73 & 1.44 & 0.87 & 1.26 \\[0.2cm]  \cline{3-9}
48 & EO Vel &&&&&&&\\
& & 2 &  2.77 & 3.28 & 1.73 & 1.66 & 0.95 & 1.58 \\
& & & & & & & &  \\
\hline
& & & & & & & &  \\
& &1 & 10.8 & 6.10 & 3.25 & 1.66 & 0.81 & 1.34 \\[0.2cm]  \cline{3-9}
49 & $\alpha$ Vir &&&&&&&\\
& & 2& 6.8 & 4.39 & 3.25 & 1.22 & 1.06 & 1.30 \\
& & & & & & & &  \\
\hline
& & & & & & & &  \\
& &1 & 13.2 & 4.81 & 4.79 & 0.83 & 1.06 & 0.91 \\[0.2cm]  \cline{3-9}
50 & DR Vul    &&&&&&&\\
& & 2& 12.1 & 4.37 & 4.79 & 0.83 & 1.12 & 0.93 \\
& & & & & & & &  \\
\hline\hline
%& & & & & & & &  \\

\end{tabular}}}

\end{table}

\clearpage

%\markboth{References}{References}
%\addcontentsline{toc}{chapter}{Литература}

\end{document}